\documentclass{ws-procs975x65}

\begin{document}

\title{A BRIEF HISTORY OF THE ENERGY-MOMENTUM TENSOR; 1900-1912}

\author{J.-P. PROVOST}

\address{Emeritus Professor, University of Nice-Sophia Antipolis, France\\
$^*$E-mail: provost@unice.fr}

\begin{abstract}
\textit{To appear in the 13th Marcel Grossmann  Proceedings, Stockholm $1^\mathrm{st}-7^\mathrm{th}$ July} 2012 \\
A critical look at the history of relativistic dynamics.
\end{abstract}

%\keywords{Style file; \LaTeX; Proceedings; World Scientific Publishing.}

\bodymatter
\bigskip

\par 1912, 100 years ago, is a remarkable date for relativity and gravitation because it marks the end of Relativistic Dynamics (R.D) and the beginning of General Relativity (G.R), both involving the energy momentum tensor ($T^{\mu \nu}$) in a crucial way. When turning from a scalar to a tensor theory, Einstein acknowledges that ``\textit{the general validity of the conservation laws} [(C.L.) $\partial_{\mu} T^{\mu \nu}=0$ or $f^{\nu}$] \textit{and the law of inertia} [$T^{i0}=T^{0i}$] \textit{is the most important new advance in the theory of relativity \ldots The problem to be solved always consists of finding how $T^{\mu \nu}$ is to be found from the variables characterizing the processes under consideration}". As discussed in Ref.~\refcite{Provost} this will explain in a large part his bumpy road to G.R between 1913 and 1915. 
\par 1900 is the time all ingredients of the future e.m $T^{\mu \nu}$ are known. Certainly Maxwell has already explained in his 1873 Treatise ``\textit{the forces} [$f=\rho E + j\wedge B$] \textit{which act on an element of a body placed in an e.m field ``by" the hypothesis of a medium in a state of stress"} [$f_i=-\partial_i(e\delta_{ij}-E_iE_j-B_iB_j)$ with $e=(E^2+B^2)/2$], and Poynting in 1884 has introduced the energy current $E  \wedge B$. In Lorentz 1900 \textit{Jubilee} Poincar\'e gives a final touch to the mechanical properties of the e.m field, providing it with a momentum density $E  \wedge B$ (in order to satisfy the action reaction principle with the charges), and submitting it to a new relativity (Galilean change $x'=x-Vt$ completed by Lorentz proper time $t'=t-Vx$ interpreted as a convention preserving the velocity of light at first order in $V$). In the same \textit{Jubilee} Wien proposes an ``\textit{e.m foundation of mechanics}" first developed by Abraham where the energy $E$, momentum $p$ and Lagrangian $L$ of an electrostatic spherical electron in global motion are deduced from the spatial integration of $e$, $E  \wedge B$ and $(B^2-E^2)/2$ in the quasistationnary approximation. This approach, unavoidable for theoreticians aware since about 20 years of the inertia of electricity (Maxwell, Thomson, ...) and suspecting a link between energy and inertia, has been fruitful because it has led to 1905 relativity and to fundamental questions in physics. For Abraham in 1902-1903, the moving electron remains spherical and for Lorentz in 1904 (who has developed his theory of ``\textit{corresponding states}" for e.m at any order in $V$), it is contracted along the direction of motion; but for both, Hamilton eq. $p=\partial L/\partial v$ is violated (in Lorentz model $p=(4/3)\gamma E_0v$, $L=-E_0\sqrt{1-v^2}$, $E=E_0\gamma(1+v^2/3)$ where $E_0$ is the electrostatic energy at rest). This problem which is related to the non trivial question of the assimilation of extended systems to punctual ones will be clarified by von Laue in 1911 (true end of R.D). It is not by chance that the citation histogram of  Einstein's 1905 relativity papers presents a bump between 1907 and 1913. 
\par In 1905, Poincar\'e writes his \textit{Palermo memoir} (59 pages, published in January 1906), which exhibits the mathematical essence (``\textit{les rapports vrais}") of Lorentz work. In particular: (i) he verifies explicitly that Lorentz eq. of dynamics $F=d(m\gamma v)/dt$ is covariant with respect to Lorentz group which he introduces and studies (this allows him to propose many relativistic gravitational forces and to attack the Mercury perihelion problem in 1906); (ii) he shows that  $L=-E_0\sqrt{1-v^2}$ arises from the invariance of the action and is linked to contraction (in short, for us today: $S=\int \phi d^3rdt =\int \phi_0 d^3r_0dt_0\longrightarrow \phi=\phi_0 \longrightarrow L$); (iii) he attributes the failure of Hamilton eq. to the instability of Lorentz electron and cures it by adding $\delta L=-Ar_0^3\sqrt{1-v^2}$ with $A=E_0/3$ (consequence of the minimization of $E_0(r)+Ar^3$ with $E_0(r) \propto r^{-1}$), but he ignores the contribution of $A$ to the mass. This contrasts with Einstein who hesitates in June concerning the eq. of dynamics but argues in September that $\Delta m=\Delta E_0$ if a body emits opposite plane waves. In March 1906, using Einstein's results on acceleration, Planck shows that Newton law at rest $ma_0=qE_0$ more generally reads $\gamma ma=q(E+v\wedge B -v(vE))$ or ``\textit{to put it in simpler form} $d(\gamma m v)/dt=q(E+v\wedge B)=F$"; he deduces the Lagrangian and the energy up to constants, i.e. ignoring the explicit covariance of the new dynamics. 
\par If in 1906 Planck criticizes the consideration of extended systems necessitating to estimate ``\textit{the work of deformation}", he soon comes back to them. Aware that one can no longer separate kinetic from internal energy because radiation is omnipresent in matter, he asks Mosengeil (dead at 22 years in September 1906) to reconsider Hasenh\"orl theory of a moving black body under the light of Einstein relativity. Remarkable advances of his 1907 paper are the invariance of entropy, of the ``\textit{number of action elements} [$h$] \textit{in nature}" and the mass defect $\Delta(E_0+p_0V_0)$ in chemistry. Technically, he uses Mosengeil results to deduce the entropy $S=\gamma^4VT^3$ and Helmholtz energy $H=vp-K+TS$ which enters the Least Action Principle $\int H dt$. A lot of tedious calculations (among which the  transformation law of the force $F=d(\partial H/\partial v)/dt$) leads him to the invariance of $\gamma (H-Cste)$; Mechanics is recovered for $Cste=0,\hspace{0.1in} S=0$. In his December 1907 review paper, Einstein considers the possibility that energy and momentum are provided to an extended moving system of charges by an applied field, under the condition that the system acquires no momentum at rest: 
$
\int dE=\int dt\int \rho vE_ad^3r; \int dp=\int dt \int \rho (E_a+v \wedge B_a) d^3r; \int dp_0=\int dt_0 \int f_0 d^3r_0=0.
$
The transformation of e.m quantities leads him by integration (and up to constants inside $E_0$) to 
$
\int dE=\gamma \int dE_0; \int dp=\gamma [\int dE_0+\int f_0 d^3r_0dt_0] \text{  i.e  if } f_0=0 \text{ to }
$
\begin{equation}
 E=\gamma (m+E_0),\hspace{0.5in} p=vE,
\label{jpp:eq1}
\end{equation}
``\textit{a result of an extraordinary theoretical importance}" (equivalent role of mass and internal energy). Being an expert in simultaneity, he is the first to understand that if $\int f_0 d^3r_0=0$ at $t_0$ fixed, with $f_0 \neq 0$ (for instance the pressure force on the walls of the black body cavity), it is no longer true at $t$ fixed. He recovers in that way Planck 1907 relation $E+pV=\gamma (E_0+p_0 V_0)$ with $p=p_0$. 

\par 1908 is the year Minkowski introduces the 4d formalism, showing in particular that Newton's law $dP^{\mu}/d\tau=F^{\mu}$ reads $\partial_{\mu} T^{\mu \nu}=f^{\nu}$ with $T^{\mu \nu}=\rho_0 u^{\mu} u^{\nu}$ (free matter), and Planck brings in physics a major conceptual unification. In \textit{Remarks on the action and reaction principle in general dynamics} he notes that energy is both various (kinetic, gravitational, calorific, chemical, e.m \ldots) and unique (through its C.L) whereas momentum is known only for mechanics and e.m. Through several examples, he shows that the energy current is nothing but the momentum density (law of inertia $T^{i0}=T^{0i}$ generalizing  Eq. (\ref{jpp:eq1})). He also claims that the stress tensor, e.g. Maxwell's one, is a momentum current which must be examined for gravitation. The physical consequences of this unification and the various expressions of $T^{\mu \nu}$ for e.m. media (Minkowski, Abraham \ldots), hydrodynamics and elasticity (Born, Herglotz \ldots) will be developed between 1908 and 1911 (see  Ref.~\refcite{Laue,Pauli}).

\par The synthesis of point particles and continuous media mechanics is made by von Laue in 1911. He shows that the above odd results are simple consequences (provided $T^{0i}_{\mathrm{rest}}=0$) of the law of transformation of $T^{\mu \nu}$: 
\begin{equation}
T^{00}=\gamma^2(T^{00}+v^2T^{11})_{\mathrm{rest}};\hspace{0.1in} T^{01}=v\gamma^2(T^{11}+T^{00})_{\mathrm{rest}};\hspace{0.1in} T^{11}-vT^{10}=T^{11}_{\mathrm{rest}}.
\end{equation}
He also gives a sufficient guarantee for $\int T^{0 \mu}d^3r$ to be a quadrivector $P^\mu$, namely that the system be static at rest and that $\int T^{ij}_{\mathrm{rest}}d^3r_0=0$. This implies $\partial_\mu T^{\mu \nu}=0$ which is clearly not satisfied for a static electron or gas; in 1918 Klein will prove that the reciprocal is true (H. Ohanian private communication). In addition, many previous ``paradoxes" such as Ehrenfest 1909 paradox (a body with $T^{12}_{\mathrm{rest}} \neq 0$ gets $p_y \neq 0$ if it is boosted along $x$), or concerning open systems such as Trouton-Noble 1903 condensator or Lewis-Tolman 1909 lever, which are submitted at rest to a zero torque but not when boosted, making them rotate if one forgot that internal stresses lead to an energy flow, i.e. to a momentum density in parts of the system. 
\par In conclusion and although it is not the standard way to look at it, the history of R.D has been implicitly or explicitly concerned with issues relative to $T^{\mu \nu}$:  integrations of e.m. densities and their insufficiency, Planck 1908 formulation of Einstein 1905-1907 inertiae, importance of its transformation law and its C.L. This is the reason why in his 1911 book on relativity \cite{Laue}, the first one on this subject, von Laue presents its chapter 7 Dynamics as ``\textit{a new exposure studying the influence of elastic stresses on energy momentum and their transformation laws}" (whereas today R.D can be introduced and developed in a few lines for students). Is the story of $T^{\mu \nu}$ finished in 1912? Of course not; from 1912 to 1921, G.R has been concerned with many issues dealing with this tensor, in particular with the status of its gravitational part in relation to the new extended covariance of the theory. Still today we ignore the status of the dark energy tensor.


\begin{thebibliography}{9}
\bibitem{Provost}  J.~P. PROVOST and C. BRACCO, in {Gravitation Quantique} (Paris, Herman, 2013)  and references therein. 

\bibitem{Laue}  M.~v. LAUE, {\em Principle of relativity} (Leipzig, 1911).

\bibitem{Pauli}  W. PAULI, {\em Theory of relativity} (Leipzig, 1921) and references therein.

\end{thebibliography}
\end{document}